\documentclass[]{aastex631}

\begin{document}

\title{Probing coronal mass ejections inclination effects with EUHFORIA}

\author[0000-0002-9866-0458
]{Karmen Martinic}
\altaffiliation{These authors contributed equally to this work.}
\affiliation{Hvar Observatory, Faculty of Geodesy, University of Zagreb \\
Kačićeva ulica 26 \\
10 000 Zagreb, Croatia}

\author[0000-0002-6998-7224]{Eleanna Asvestari}
\altaffiliation{These authors contributed equally to this work.}
\affiliation{Faculty of Science, Department of Physics, University of Helsinki \\
 Gustaf Hällströmin katu 2 \\
 Helsinki, Finland}

\author[0000-0002-8680-8267]{Mateja Dumbović}
\affiliation{Hvar Observatory, Faculty of Geodesy, University of Zagreb \\
Kačićeva ulica 26 \\
10 000 Zagreb, Croatia}

\author[0000-0003-2617-4319]{Tobias Rindlisbacher}
\affiliation{Albert Einstein Center for Fundamental Physics, Institute for Theoretical Physics, University of Bern, \\
Sidlerstrasse 5, CH-3012 \\
Bern, Switzerland}

\author[0000-0003-4867-7558]{Manuela Temmer}
\affiliation{Institute of Physics, University of Graz \\
Universitätsplatz 5 \\
8010 Graz, Austria}

\author[0000-0002-0248-4681]{Bojan Vršnak}
\affiliation{Hvar Observatory, Faculty of Geodesy, University of Zagreb \\
Kačićeva ulica 26 \\
10 000 Zagreb, Croatia}



\begin{abstract}

Coronal mass ejections (CMEs) are complex magnetized plasma structures in which the magnetic field spirals around a central axis, forming what is known as a flux rope (FR). The central FR axis can be oriented at any angle to the ecliptic. Throughout its journey, a CME will encounter interplanetary magnetic field and solar wind which are neither homogeneous nor isotropic. Consequently, CMEs with different orientations will encounter different ambient medium conditions and, thus, the interaction of a CME with its surrounding environment will vary depending on the orientation of its FR axis, among other factors. This study aims to understand the effect of inclination on CME propagation. We performed simulations with the EUHFORIA 3D magnetohydrodynamic model. This study focuses on two CMEs modelled as spheromaks with nearly identical properties, differing only by their inclination. We show the effects of CME orientation on sheath evolution, MHD drag, and non-radial flows by analyzing the model data from a swarm of 81 virtual spacecraft scattered across the inner heliospheric. We have found that the sheath duration increases with radial distance from the Sun and that the rate of increase is greater on the flanks of the CME. Non-radial flows within the studied sheath region appear larger outside the ecliptic plane, indicating a "sliding" of the IMF in the out-of ecliptic plane. We found that the calculated drag parameter does not remain constant with radial distance and that the inclination dependence of the drag parameter can not be resolved with our numerical setup.

\end{abstract}

\keywords{Solar coronal mass ejections (310); Magnetohydrodynamical simulations
(1966); Space weather (2037); Interplanetary magnetic fields (824)}


\section{Introduction} \label{sec:intro}

Coronal mass ejections (CMEs) are massive expulsions of plasma and magnetic fields from the solar corona into interplanetary space (IPS). While traveling through the IPS, CMEs are also referred to as Interplanetary CMEs (ICMEs) and are known for their potential geoeffectiveness (\citealp{Gosling1990}, \citealp{Zhang2003} \citealp{Koskinen2006}, \citealp{Dumbovic2015}). The magnetic core of (I)CMEs is known to be a twisted flux rope (FR) (\citealp{Burlaga1981}, \citealp{Lepping1990}, \citealp{Vourlidas2013}), namely a cylindrical structure whose poloidal magnetic field component wraps around an axial magnetic field that follows the central axis of the cylinder \citep{LUNDQUIST1950}. 

During their propagation in the corona and the heliosphere, (I)CMEs interact with the surrounding solar wind (SW) and the interplanetary magnetic field (IMF). As a result of this interaction, and in combination with the FR expansion, a so-called sheath region is formed ahead of the FR \citep{Siscoe2008}. This highly compressed and turbulent part of the (I)CME usually exhibits increased magnetic field, temperature, and density, which has strong fluctuations \citep{Kilpua2017}. (I)CME sheath regions are known for their geoeffectiveness (\citealp{Tsurutani1988}, \citealp{Huttunen2002}, \citealp{Huttunen2004}) and are therefore very important to study for advancing space weather forecasts. 

Sheath regions consist of draped IMF and accumulated SW plasma in front of the FR driver. \cite{Gosling1987} introduced a concept of IMF draping around (I)CMEs in IPS, analogous to the draping of IMF around planets and comets forming various plasma sheaths. They developed a the theory based on the draping of the radial and transverse IMF components. In the case of the radial one, it is related to the Gosling-McComas rule according to which the amplification of the negative $B_z$ perturbations occurs in the (I)CME sheath region. There are two cases when negative $B_z$ occurs: 1) in front of a northward directed (I)CME that is embedded in sunward (inward pointing - negative) radial IMF (see Figure 3 in \citealp{Gosling1987}) and 2) in front of the southward directed (I)CME that is embedded in anti-sunward (outward pointing - positive) IMF. \cite{McComas1989} reported a strong out-of-ecliptic magnetic field component in the sheath region of an observed (I)CME and discussed that this is a consequence of a draped magnetic field pattern following the deflection of the surrounding plasma in front of the FR driver. However, the draping is a complex 3-dimensional (3D) pattern in which both radial and transverse magnetic field components are draped around the embedded (I)CME. The draping of the transverse magnetic field component leads to two additional mechanisms: the eastward deflection of fast (I)CMEs \citep{Gosling1987b} and $B_z$ perturbations due to the pressure gradient that shifts the accumulated field lines perpendicular to the direction of motion of the (I)CME. \cite{Siscoe2007} employed simulations to analyze the draping of the transverse IMF component for fast CMEs. They found that the magnetic field is weaker on the eastern side and stronger on the western side of the (I)CME front. They also concluded that, at higher latitudes, the latitudinal component of the magnetic field is stronger on the eastern side. This is a consequence of the existence of the Parker spiral, i.e. the draping of the transverse magnetic field component, where a stronger draping of the IMF occurs on the western side and a "slipping“ of the field lines on the eastern side of the (I)CME front is a more probable process.

More recently, \cite{Martinic2022} and \cite{Martinic2023}, studied different drag force, and the different plasma outflows (i.e. the different draping patterns) experienced by CMEs with different inclinations (CMEs inside and outside the ecliptic plane), assuming that IMF is "frozen-in" the SW plasma. They found observational evidence for differences in the non-radial flows in the CME sheath region for differently inclined CMEs. However, no evidence was found for a difference in drag for CMEs of different inclinations. \cite{Vandas1995} and \cite{Vandas1996} performed magnetohydrodynamic (MHD) simulations of two differently inclined magnetic clouds. Details of the MHD model used can be found in \cite{Wu1979} (for 2D simulations) and in \cite{Wu1983} (for 2.5D simulations). Their results show that the trajectory of these magnetic clouds is unaffected by the alignment of their axes with the ecliptic plane, regardless of whether one of the two axes lies within the ecliptic and the other has an axis perpendicular to it. However, it is important to note that the MHD model they used was limited to the equatorial plane of the Sun and did not provide a comprehensive 3D MHD representation.

This study aims to fill the gap in understanding the effects of inclination on propagation within the heliosphere by performing simulations using the EUropean Heliospheric FORecasting Information Asset (EUHFORIA) 3D MHD model \citep{Pomoell2018}. Considering the limitations in observing and isolating the effects of CME inclination highlighted in \cite{Martinic2022} and \cite{Martinic2023}, this research will simulate two CMEs with nearly identical properties, differing only in their inclination, using the spheromak CME representation in EUHFORIA \citep{Verbeke2019}. In section \ref{Numerical_setup} we first present the numerical setup for the simulations and discuss the determination of the spheromak boundaries in section \ref{boundary_determination}. In sections \ref{Spheromak_tilting} and \ref{Drag_parameter_determination} we discuss, respectively, the importance of the spheromak tilt instability and the drag parameter $\gamma$. The sheath evolution, non-radial flows, and the drag parameter $\gamma$ for two differently inclined spheromaks are shown and discussed in the section \ref{Results}. Finally, a summary of the main results is given in section \ref{Conclusions}.

\section{Numerical Setup}
\label{Numerical_setup}
For this study, we generated simplistic solar wind plasma and IMF conditions with EUHFORIA MHD model. At the inner boundary, set at 0.1~AU, we input a uniform, weak, and outward pointing (positive) radial IMF, of $B_{r}=$100~nT. The selection of this weak IMF was chosen to minimize the tilting and deflecting effects of the IMF on the spheromak \citep[see][for more details on these phenomena]{Asvestari2022}. As argued in \citet{Asvestari2022}, although this unidirectional IMF topology is not realistic for the global 3D heliosphere, it can be a valid topology for the small region where a FR is injected and through which it propagates. The plasma parameters at the inner boundary are set at v$_{r}$ = 400.0~km~s$^{-1}$ for the radial velocity component and P = 3.3~nPa for the thermal pressure, a representative for the average properties of a slow solar wind at the inner boundary. Based on these two parameters the number density was calculated using equation (4) in \citet{Pomoell2018} giving n = 7.32~$\cdot$~10$^8$~m$^{-3}$. Subsequently, using the number density and plasma pressure, the plasma temperature was calculated and found to be at T = 3.26~$\cdot$~10$^5$~K. For the grid resolution of the 3D domain, we set 512 radial grid points from 0.1~AU - 2.0~AU, and 4$^\circ$ resolution in co-latitude and longitude. We performed two simulations, one with a spheromak inserted with a 0$^\circ$ tilt angle (low inclination), and one at a 90$^\circ$ tilt angle (high inclination). Tilt angle is defined as an angle between the solar equatorial plane and the toroidal axis of the spheromak. Both spheromakstructures were set to have a positive helicity and were inserted at 0$^\circ$ longitude and latitude (Sun-Earth line), with a radius of 10~R$_{\odot}$. For the magnetic flux and temperature of the spheromak, we opted to use default EUHFORIA values, which are 80~$\cdot$~10$^{12}$~Wb and 0.8~$\cdot$~10$^6$~K, respectively \citep{Verbeke2019}. To further minimize the rotation of the spheromak due to the ambient magnetic field we opted for modeling the spheromak having relatively high density of $\rho$ = 0.5~$\cdot$~10$^{-17}$~kg~m$^{3}$ \citep{Temmer2021}, thus more inertia/harder to rotate, and which are faster, v = 1500.0~km~s$^{-1}$, thus escaping quickly from the inner heliosphere where the ambient field is stronger and has a greater impact \citep[see][for more details]{Asvestari2022, Sarkar2024}.

By choosing a simple, unidirectional configuration of the surrounding magnetic field, we avoid the occurrence of the heliospheric plasma sheath (HCS) in the simulation. \cite{Smith2001} and \cite{Temmer2023} emphasize the complexity and importance of studying the interaction between the HCS and the CMEs. It is known that CMEs deflect towards the HCS (\citealp{Kay+2015}; \citealp{Wang+2023}), align their axis with the local orientation of the HCS (\citealp{Yurchyshyn2008}; \citealp{Yurchyshyn2009}) and cause FR erosion through reconnection (e.g. \citealp{Dasso2006}; \citealp{Ruffenach2015}; \citealp{Pal2021}). As EUHFORIA is an ideal MHD model, we do not anticipate physical reconnection to take place. However, reconnection may occur in the simulation domain due to numerical diffusion caused by the grid resolution and/or the numerical solver. The grid resolution mainly affects the HCS, which, although it should be infinitesimal in thickness, is smeared, resulting in a numerical resistivity that can lead to reconnection. Additionally, the configuration of the ambient field (pointing radially outward) and helicity of the spheromaks (positive helicity) is chosen to minimize the occurrence of the anti-parallel magnetic field lines of the spheromak's poloidal field and the ambient magnetic field in the simulation.

\begin{figure}[ht!]
\centering
\includegraphics[width=0.4\linewidth]{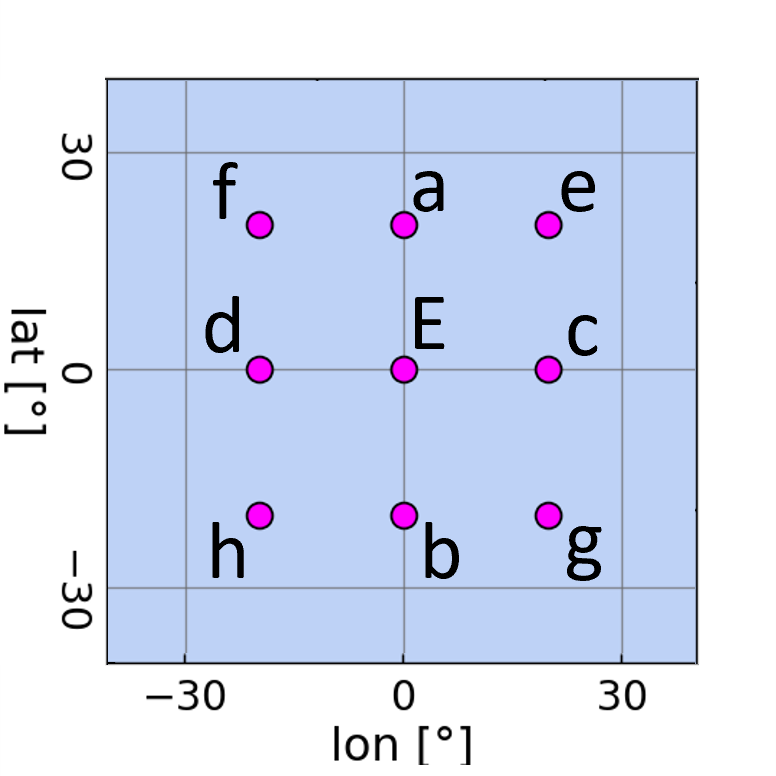}
\caption{Longitudinal and latitudinal distribution of the virtual spacecraft swarm used in EUHFORIA simulations for particular radial distance.
\label{fig1}}
\end{figure}

To study the spheromak's inclination effects on its evolution and on the draping of IMF at its sheath at different distances from the Sun and different locations at its front and flanks we inserted virtual spacecraft (VS) in the simulations.
 The spacecraft names do not reflect their location; they are just assigned to different lon-/lat-positions. Overall we inserted 81 VS, grouped into 9 families. Each family consists of 9 VS with identical longitudinal/latitudinal coordinates but having different radial distances from the Sun. The labeling scheme for the 9 VS families is shown in Figure \ref{fig1}. As can be seen in this figure we placed one VS along the Sun-Earth line (0$^\circ$ in longitude and latitude), which is noted as VS-E, and 8 VS (a-h) in a square grid around VS-E, using longitudinal and/or latitudinal separations of 20$^\circ$ (see Figure \ref{fig1}). This spacecraft constellation is repeated for radial distances of 0.32~AU, 0.42~AU, 0.52~AU, 0.62~AU, 0.72~AU, 0.92~AU, and 1.00~AU, determined by the grid resolution. The coordinate system in the simulations is in Heliocentric Earth Equatorial (HEEQ).

\section{Spheromak boundary determination}
\label{boundary_determination}

For our analysis, it was essential to determine, at each VS location, the boundaries between undisturbed solar wind, shock-sheath, and FR regions, namely, the start and end time of the sheath region and of the FR, hereafter referred to as the magnetic obstacle (MO). To do that we employed two methods; the first was based on analysing the plasma $\beta$ parameter in the time series at each VS, similar to criteria observers use with in-situ measurements. This approach is based on the fact that within the MO part of the (I)CME, the magnetic pressure dominates over the plasma pressure, while the opposite applies in the sheath region. The second method focused on analysing the 3D structure in the modelling domain. More details on this approach are given later in this section.

In the first method, the onset time of the sheath is determined by the sudden increase of the plasma $\beta$ parameter, i.e., a 50$\%$ increase of the plasma $\beta$ compared to the mean of the preceding 35 plasma $\beta$ values in the solar wind. The end of the sheath is defined as a sudden increase in the plasma $\beta$ parameter, relative to the mean of the trailing 35 plasma $\beta$ values in the solar wind, when iterating from the back of the timeseries. We note that this sheath end boundary would be seen as a sudden decrease in plasma $\beta$ parameter, if we start from the beginning of the time series This criterion for determining the sheath boundaries is based on evidence that the plasma $\beta$ parameter in the sheath region increases in (I)CMEs examined in-situ \citep[][and reference therein]{Kilpua2017}. The start of MO was determined at a point where the plasma $\beta$ falls below one, and the end at the time at which it reaches a value greater than one again. The same condition for obtaining MO boundaries based on the EUHFORIA simulation output with spheromak was applied in \cite{Sarkar2024}. 

\begin{figure}[ht!]
\plotone{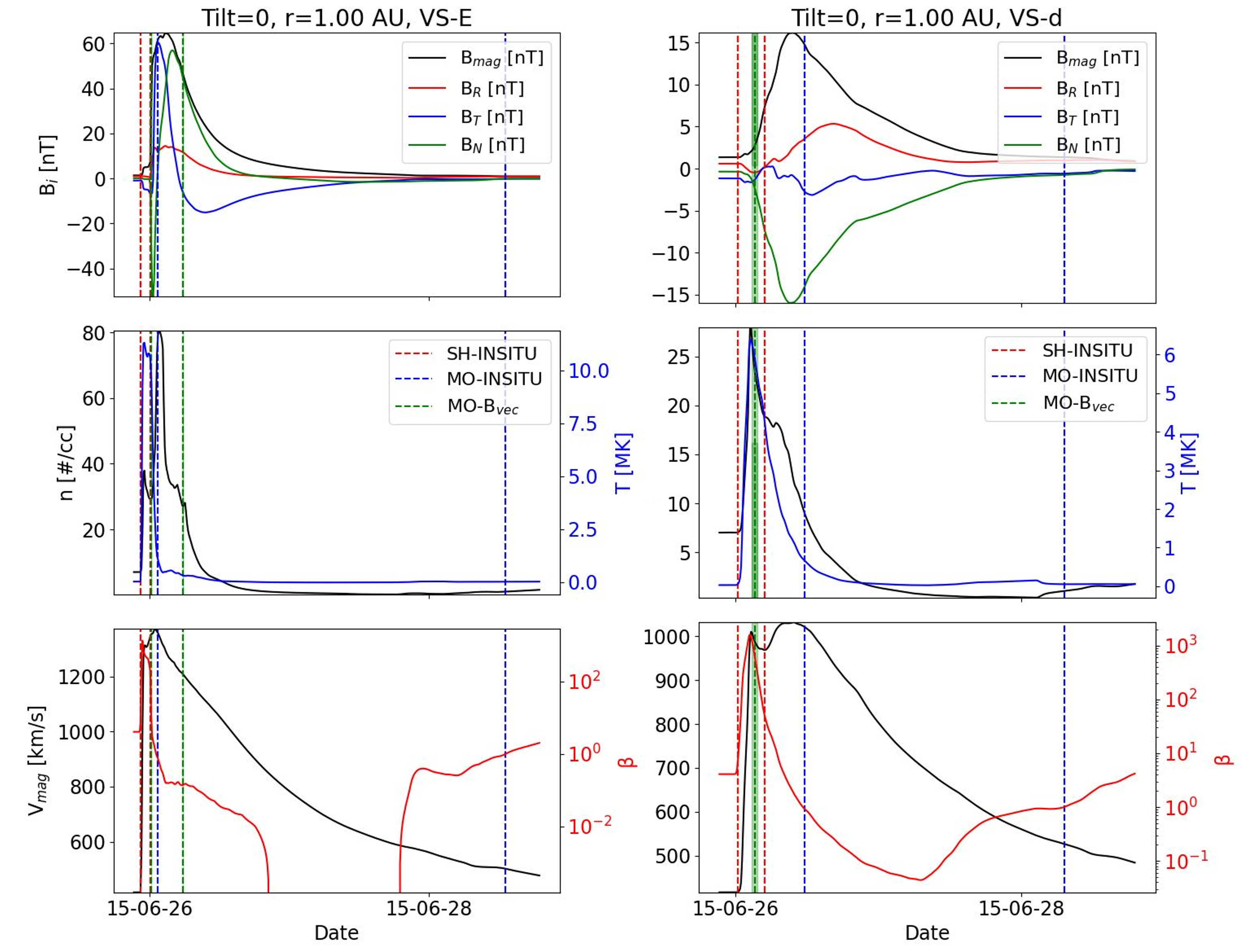}
\caption{In-situ data as seen by VS-E (left panels) and VS-d (right panels) for 1 AU distance. The top panels show magnetic field magnitude and magnetic field components in the RTN coordinate system. Middle panels show proton number density and temperature. The bottom panels show velocity and plasma $\beta$ parameters. The pair of red vertical boundaries mark the sheath region (SH-INSITU), the pair of blue vertical boundaries mark the MO part based on the plasma $\beta$ parameter (MO-INSITU), first green vertical line marks the beginning of the MO part but based on the magnetic flux density as a vector field (MO-B$_{vec}$), while the second green vertical line (only present for VS-E, left panels) marks the end of the spheromak also derived on the basis of the magnetic flux density as a vector field (MO-B$_{vec}$). 
\label{fig2}}
\end{figure}

Figure~\ref{fig2} shows the in-situ profiles for VS-E and VS-d located at $r=1$~AU for a spheromak with high inclination (tilt=90). From top to bottom, the panels display the magnetic field magnitude and its components in the radial-tangential-normal (RTN) coordinate system (top row), the proton number density and the temperature (middle row), the velocity and the plasma $\beta$ parameter (bottom row). The left column shows the values for VS-E and the right for VS-d. In addition, five (for right panels) and 6 (for left panels) vertical lines can be seen in the figure. These mark the boundaries of different (I)CME signatures. The sheath, which was derived solely on the basis of the plasma $\beta$ parameter, is marked with two red vertical lines. The two blue vertical lines mark the beginning (left line) and end (right line) of the MO, again, solely based on the plasma $\beta$ parameter. The green vertical line marks the start of the MO derived based on the second approach described below. The end of the MO, based on the second approach, is set only for VS-E because only this boundary is used in section \ref{Drag_parameter_determination} to derive the drag parameter $\gamma$.  

\begin{figure}[ht!]
\plotone{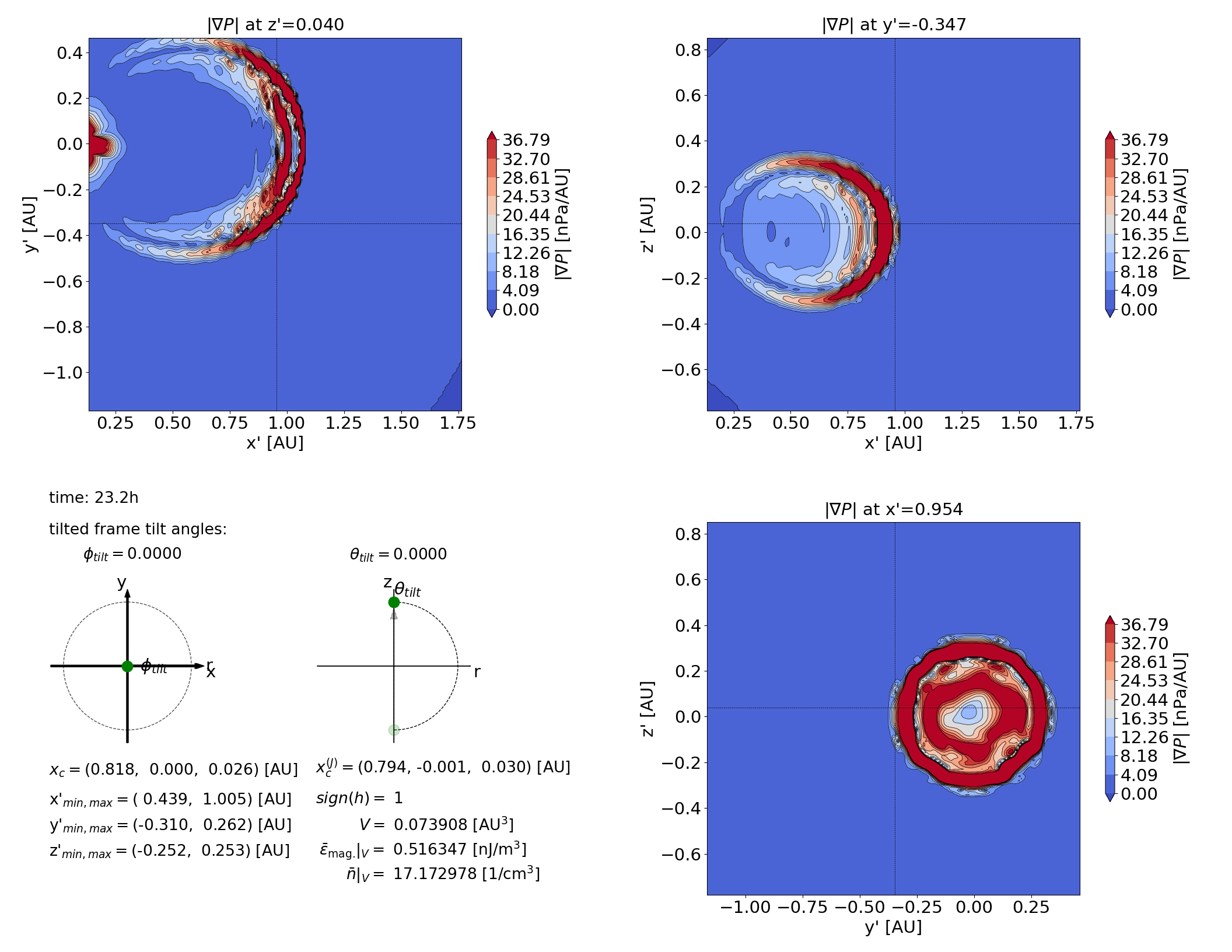}
\caption{The magnitude of the pressure gradient in a neighborhood of VS-d in the low inclination spheromak simulation, visualized along 3 perpendicular slices (two top and bottom right panels). The 3 slices are selected so that VS-d is located at their centers (grid cross). The $x'$-$y'$-$z'$ coordinate system, specified by the orientation of the slices, agrees here with HEEQ ($x$-$y$-$z$). The bottom left panel shows information about the spheromak's center location, extent, helicity, volume, average magnetic field energy, and mass, as determined by the detection tool. The angles $\phi_{tilt}$ and $\theta_{tilt}$ specify the orientation of the $x'$-$y'$-$z'$ coordinate system relative to the HEEQ $x$-$y$-$z$ coordinates.
\label{fig3}}
\end{figure}

\begin{figure}[ht!]
\plotone{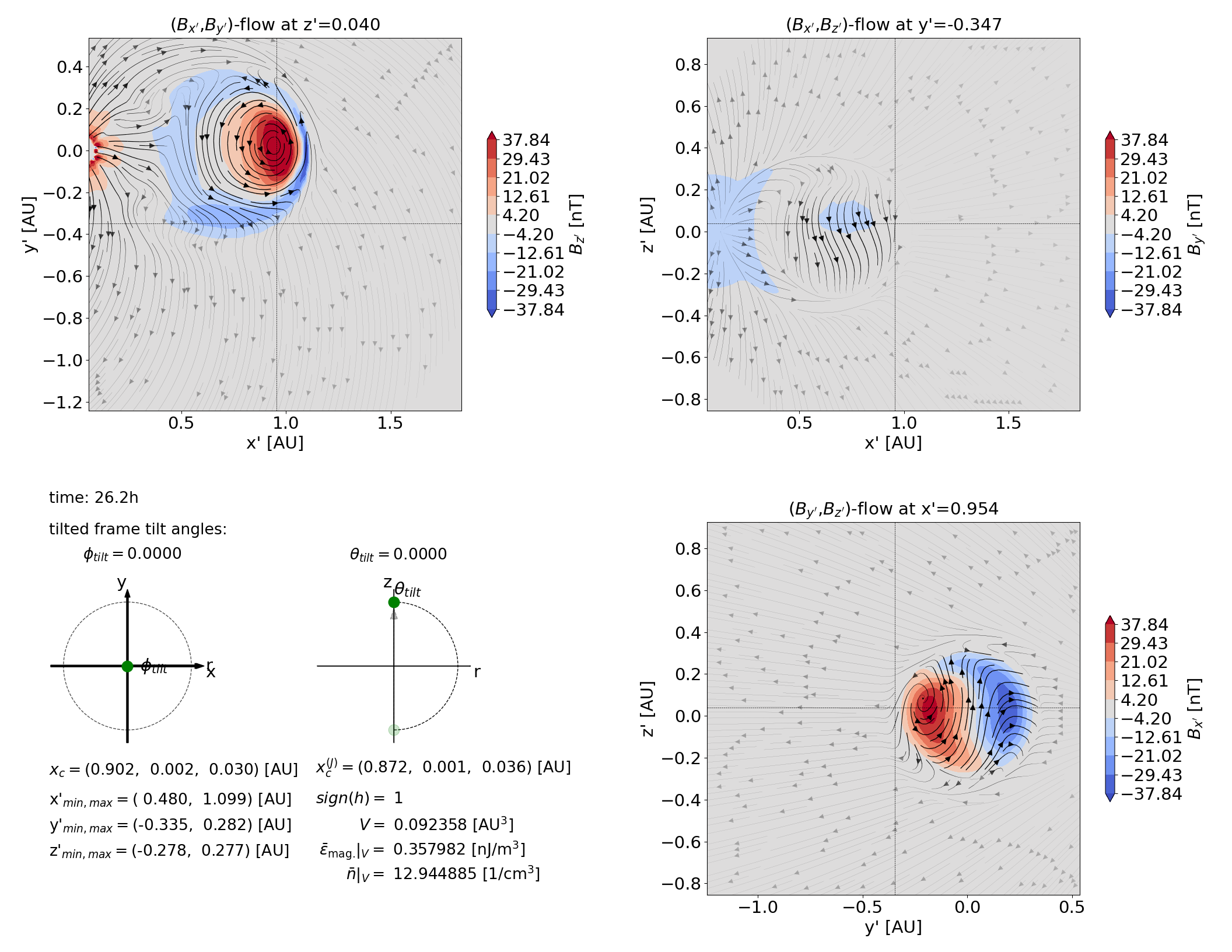}
\caption{The vector field indicating the division between spheromak FR and IMF for VS-d in the low inclination spheromak simulation. Same as the previous figure, the 3-slices are selected so that the spacecraft being analyzed is located at the center of the images (grid cross).
\label{fig4}}
\end{figure}

The second method focuses on analysing the entire spheromak structure in the 3D simulation output. These boundaries are marked by the green vertical lines in Figure~\ref{fig2} which correspond to the beginning and the end of the MO derived by analyzing the 3-dimensional spheromak structure based on the magnetic field and magnitude of the plasma pressure gradient. More precisely we employed an improved version of the spheromak detection and tracking method described in \citet{Asvestari2022}. As the different structures of the (I)CME were traversing a VS, the quantities of interest were monitored along three perpendicular cut planes, centered at the VS location and oriented to be parallel to the HEEQ coordinate planes. First, we looked into the magnitude of the pressure gradient $|\nabla P|$, an example of which is shown in Figure~\ref{fig3} for VS-d for a spheromak of low inclination. As can be seen, we can identify two thin shells of the increased magnitude of plasma pressure gradient. In between these two shells, the pressure gradient vanishes (in perpendicular direction to the shells' surfaces), since the pressure itself peaks along the interface between CME and ambient solar wind, where they push directly against each other. The start of the first shell marks the start of the sheath region, while the start of the second shell marks the beginning of the MO. Therefore, tracking the arrival of each shell at the VS location, the center of the grid cross in all panels, helps us to extract the start times. It is important to mention that the wiggles these shells exhibit are due to the grid resolution and the manifestation of numerical artifacts. In the case of the MO, we looked also at the magnetic flux density as a vector field. An example is shown in Figure~\ref{fig4} for the same VS-d as in Figure~\ref{fig3}. The moment at which the spheromak arrives at the VS is determined by picking the time at which the spheromak's magnetic field becomes visible in all three panels (which are cut planes through the VS location). The example in Figure~\ref{fig4} shows such a moment, since the spheromak's poloidal field just started to appear in the top-right panel.

Using the same plot we could determine the end of the FR. To do that we need to consider that the spheromak FR structure is "ring-shaped" which means that the leading FR is followed by a trailing one with mirrored poloidal component and anti-parallel toroidal one. This double FR structure is known to not represent the commonly accepted geometry of the MO of (I)CMEs and thus we only want to consider the leading FR in our analysis. Consequently, to determine the end of the MO we consider the time the spacecraft crosses the poloidal field for the second time. The end of the MO was derived with a higher temporal resolution of the 3D simulation output, 5 min, but only for VS-E, as shown in Figure~\ref{fig2}, left panels. The end of the spheromak was needed only for VS-E to derive the drag parameter $\gamma$ as described in section \ref{Drag_parameter_determination}, and thus for other VS the end of the MO is not determined. The green shaded area around the green vertical line marking the beginning of the MO (for all VS, except for VS-E for which higher resolution simulation output was used) in Figure~\ref{fig2} represents the temporal resolution of the images we used to derive the beginning of the MO, i.e. it corresponds to $\pm$~30 min from the set boundary time as shown in the right panels of Figure \ref{fig2}. To summarize, using the second method, the beginning of the MO was derived for all VS, while the end of the MO was derived only for VS-E. Also, for VS-E MO boundaries based on the second method were derived with higher simulation output cadence- 5 min.

In Figure \ref{fig2} (left panels) we can also see that the boundaries of the end of the MO, derived by the two different methods explained, do not coincide. This is a direct consequence of the spheromak's toroidal geometry. The end of the MO with the green vertical line is located in the center of the toroid of the spheromak and the rear part of the toroidal magnetic body is not considered, which is not the case when estimating the boundary of the MO solely based on the plasma $\beta$ parameter in in-situ time series (second blue vertical line). Note that in the simulations the spheromak does not retain its spherical geometry. Instead, it expands and pancakes as it interacts with the ambient solar wind, resulting in a compressed front and a sun-ward stretched trailing portion. This explains why the green line marking the end of the MO in Figure~\ref{fig2} (left panels) is close to the front and not at the mid-point of the spheromak FR signature in the time series.

The in-situ profiles with all corresponding boundaries and for all spacecraft (overall the different radial distances in the simulation), for high and low-inclination spheromak, can be found at the following link: \href{https://doi.org/10.6084/m9.figshare.25849135}{doi:10.6084/m9.figshare.25849135}. We point out that for VS-a and VS-e, in the case of the high inclination spheromak, and for VS-c and VS-g, in the case of the low inclination spheromak, the end of the sheath (second red vertical line) has been manually derived due to simulation artifacts occurring just after the MO part of the spheromak. Furthermore, we note that for some spacecraft, the MO boundaries based on the $\beta$ parameter of the plasma (blue vertical lines) are not set. This is because, in some in-situ profiles, there is no plasma $\beta$ less than one. Details on the manually derived end of the sheath and the examples of in-situ profiles where plasma $\beta$ does not fall below one are shown in Appendix \ref{boundary_determination_apendix}. Plots of the plasma $\beta$ parameter across all radial distances with corresponding MO and sheath boundaries overplotted for both, high and low inclination spheromak are given in the following link: \href{https://doi.org/10.6084/m9.figshare.25849135}{doi:10.6084/m9.figshare.25849135}. 

\section{Manifestation of spheromak tilting and drifting}
\label{Spheromak_tilting}
\begin{figure}[ht!]
\includegraphics[width=\linewidth]{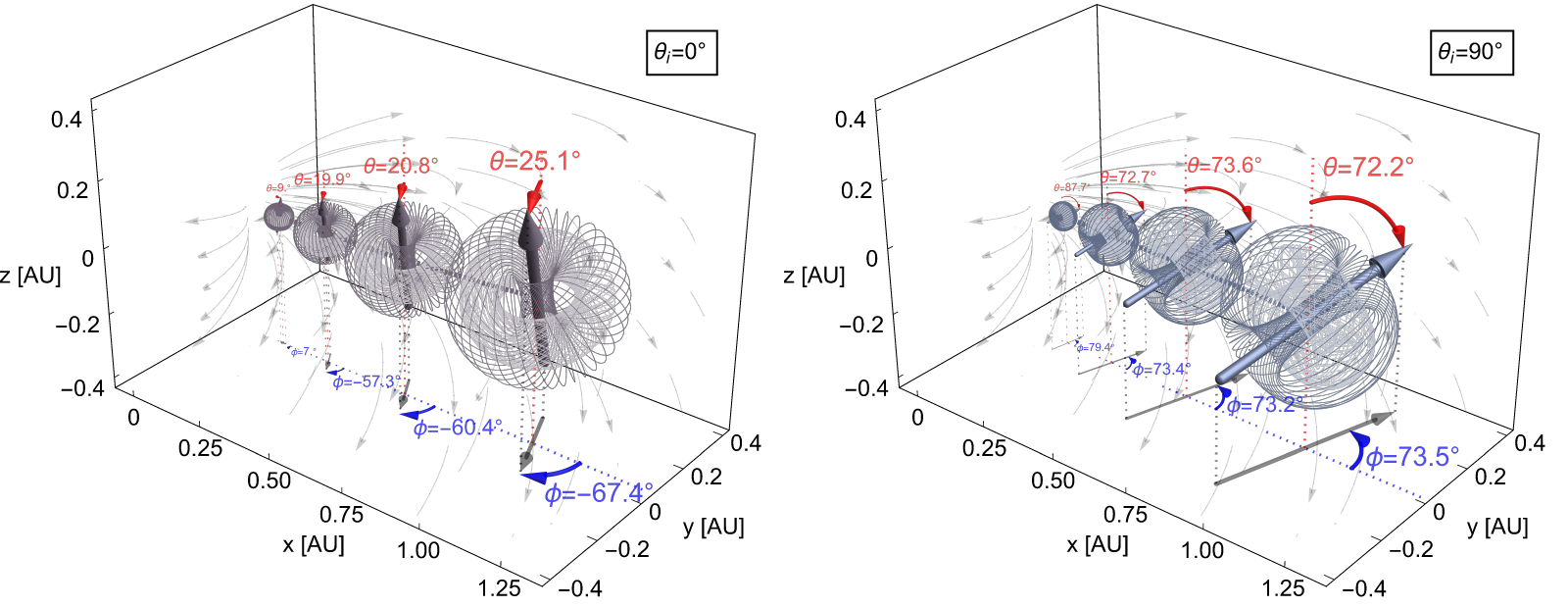}
\centering
\caption{Visual representation of how the spheromak of low inclination (left panel) and high inclination (right panel) has tilted from its original orientation due to the torque exerted on it by the ambient magnetic field. For simplicity we sketch the spheromaks in their idealized representation and at only 80\% of their actual size to avoid overlap. The arrow running through the center of the spheromak represents its magnetic moment, indicating the orientation of the poloidal field at the center of the spheromak. The change of orientation of this arrow illustrates the spheromak tilting towards the IMF-direction (gray arrows). This tilting is quantified by two angles, $\theta$ (red) and $\phi$ (blue), where $\theta$ is the angle the magnetic moment forms with the z-axis and $\phi$ is the angel the projection of the magnetic moment on the x-y-plane forms with the x-axis. The evolution of these angles over the full simulation time can be found in Figure~\ref{fig6} below.}
\label{fig5}
\end{figure}

\begin{figure}[ht!]
\includegraphics[scale=0.7]{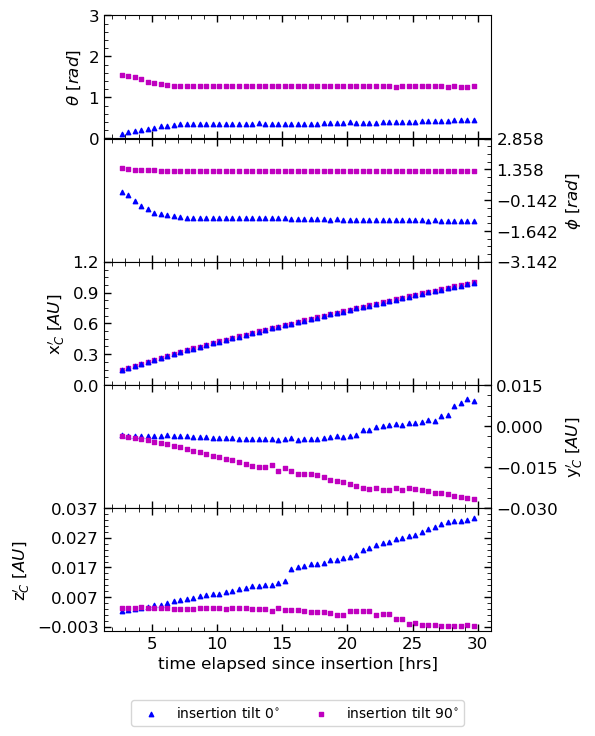}
\centering
\caption{The evolution of the spheromak orientation due to the torque causing it to tilt (first and second panels from the top), and its drifting motion indicated by the change of the location of its center of mass in the $x'$, $y'$, and $z'$-coordinates (third-fifth panels from the top). The first panel from the top shows the spheromak tilting from the solar north towards the solar equatorial plane ($\theta_{tilt}$), while the second panel shows the tilt in the solar equatorial (x--y) plane $\phi_{tilt}$. 
\label{fig6}}
\end{figure}

In \citet{Asvestari2022} it was shown that, when inserted into the interplanetary magnetic field, the spheromak can start rotating (tilting) due to a torque forcing it to align its magnetic moment to the direction of the ambient magnetic field, which in the heliospheric MHD simulations is the IMF. Furthermore, the spheromak drifts away from its initial direction of propagation. To minimize this rotation and drift, we opted, as mentioned in Section~\ref{Numerical_setup}, for inserting rather heavy spheromaks in the simulation domain, with mass densities of 0.5~$\cdot$~10$^{-17}$~kg~m$^{3}$. This is a similar approach to \citet{Sarkar2024}. The reasoning behind this choice is that a heavier spheromak has larger moment of inertia than a lighter spheromak of the same size, and will therefore experience less rotational acceleration when subject to the same torque. Similarly, more mass means more inertia and therefore less acceleration/deflection when subject to the same force. Of course, the rotation is not completely prevented but heavily reduced. In the cases we studied, the evolution of the orientation of the two spheromaks simulated with EUHFORIA, with low and high inclination, is shown, respectively, in the left- and right-hand panel of Figure~\ref{fig5}. For simplicity, we sketch the spheromak in their idealized representation. The arrow running through the center of the spheromak is its magnetic moment, indicating the orientation of the poloidal field at the center of the spheromak. As the spheromak tilts due to the torque exert on it by the IMF, the orientation of this arrow changes accordingly. As can be seen in the left-hand panel, for the low inclination spheromak, there is a slight eastward and southward tilting, while for the high inclination spheromak in the right-hand panel, we see a slight northward and very small, basically negligible, westward tilting. This is because for both cases the IMF, depicted as swarm of gray arrows, is positive and thus points away from the Sun, while at the same time it curves forming the Parker Spiral. The tilt and the drift of the two spheromaks modeled can also be seen in the time series presented in Figure~\ref{fig6}. 

The two upper panels show the time evolution of spheromak tilt, represented by the two angles $\theta$ and $\phi$ (cf. bottom-left panel of Figure~\ref{fig9}). As can be seen, the overall change in $\theta$ is for both spheromak relatively small and occurs primarily during the first 5-7 hours, while propagating below 0.3~AU. After that distance, the variations become almost unnoticeable. The third to fifth panels show, respectively, the time evolution of the x, y, and z coordinates of the spheromaks' centers of mass. As can be seen, the drift in the y- and z-directions remains for both spheromak very small, of the order of $~10^{-2}$~AU.
This small-scale drift in the y- and z-direction is of the opposite direction for different inclination spheromaks, low inclination spheromak drifts westward, and high inclination spheromak drifts eastward (see 4th panel of Figure~\ref{fig6}). Similarly, as can be seen from the bottom panel of Figure~\ref{fig6} the low inclination spheromak drifts slightly northward, and the high inclination spheromak drifts slightly southward, but such southward drift is almost negligible. 

\section{Drag parameter $\gamma$ determination}
\label{Drag_parameter_determination}

Studies based on observations have revealed that the drag force becomes the dominant factor governing the propagation of (I)CMEs within specific distances in the heliosphere. Several drag-based models for (I)CMEs have been established based on these findings (\citealp{Vrsnak2013}, \citealp{Hess2015}, \citealp{Mostl2015}, \citealp{Kay2013}). These models commonly utilize a simple analytical equation:

\begin{equation}
a_d=\gamma (v-w)|v-w|,
\end{equation}

\noindent where $v$ denotes the CME velocity, $w$ represents the solar wind velocity and $\gamma$ stands for the drag parameter calculated using the equation \citep[e.g.][]{Vrsnak2013}:

\begin{equation}
\gamma=\frac{C_d}{L(\rho/\rho_w + 1/2)}.
\end{equation}

\noindent Here, $L$ corresponds to the radial cross-section of the (I)CME MO part, $\rho$ is (I)CME MO enclosed density, $\rho_w$ the ambient solar wind density, and $C_d$ the dimensionless drag coefficient. Notably, $C_d$ is often assumed to be constant and set to one throughout the (I)CME propagation. However, \cite{Cargill2004} demonstrated that the value of $C_d$ depends on the relative density and velocity of the (I)CME with regard to the solar wind. Cargill's work highlighted that for dense CMEs, $C_d$ remains close to one, whereas for low-density (I)CMEs, it can increase to as high as three, exhibiting significant radial dependence in the latter case.

The drag parameter $\gamma$ plays a crucial role in understanding the drag force acting on an (I)CME. From the EUHFORIA simulation, we can derive the density ratio $\rho/\rho_w$ for each VS at each radial distance from the Sun in the heliosphere. The radial cross-section, $L$, can also be derived from simulation results. From in-situ representation, $L$ is given as the duration of the MO multiplied by the mean plasma velocity inside the MO as observed by the VS. We note here that the above-given representation of the drag parameter $\gamma$ assumes locally cylindrical geometry where $L$ is two times the radius of the base of the cylinder (for details see \citealp{Vrsnak2013} and \citealp{Cargill2004}), and to be the closest to this idealization we only calculated the cross-section $L$ from green MO boundaries displayed in Figure~\ref{fig2}. This way we avoid a double FR crossing due to the spheromak "ring-shaped" toroidal magnetic body. The parameter $\rho$ is taken as a median of the density measurements at the VS and $\rho_{w}$ is the mean density before the (I)CME sheath onset (mean ambient density). Finally, $C_d$ is assumed to be constant and is set to one.

\section{Results and discussion}
\label{Results}

\subsection{Evolution of the Sheath region}

Figure~\ref{fig2} shows that for VS-E the end of the sheath (second red vertical line) and the start of the MO (first green vertical line) coincide. However, this is not the case for VS-d. We see that in this case, the MO start (first vertical green line) is before the estimated end of the sheath (second red vertical line). As a reminder for the reader, the end of the sheath was based solely on the plasma $\beta$ while the start of the MO was derived based on the magnetic field vector flow. Therefore, the fact that the start of the MO was found to be before the determined end of the sheath means that the plasma $\beta$ parameter stays increased within the beginning of the MO. Within this region (between the first green vertical line and the second red vertical line), as shown in Figure~\ref{fig2} (right panels) we see a gradual decrease in velocity, while temperature and number density show a rapidly decreasing profile similar to plasma $\beta$ parameter. \cite{Vrsnak2016} studied early FR evolution in the solar corona in the 2.5 D MHD setting and found a sharp density peak in front of the FR represents the contact surface, i.e., the edge of the
FR where the density pile-up forms as a result of the flux-rope expansion. From an observational point of view, the compressed region in front of the MO part of (I)CMEs, which has both sheath and MO properties, has been studied in \cite{Kilpua2013} and in \cite{Temmer2022}. Interestingly, this is also manifested in our simulation setting. However, it is beyond the scope of this work to determine the origin of the misalignment between the sheath end time and MO onset time for some of the VS and to relate simulation results with observational features of this region. From now on, we will call this region the \textit{MO front} region for simplicity. For the purpose of this study, we first looked into the extended sheath signatures, where the duration of the sheath is given as $SH_{end}$-$SH_{start}$ (region in-between red vertical lines). We next analyze the clear sheath region with duration $MO_{start}$-$SH_{start}$ (region in-between first red and first green vertical line). Finally, we analyze the MO front region with duration $MO_{start}$-$SH_{end}$ (between the first green vertical line and the second red vertical line).

\begin{figure}[ht!]
\centering
\includegraphics[width=1\linewidth]{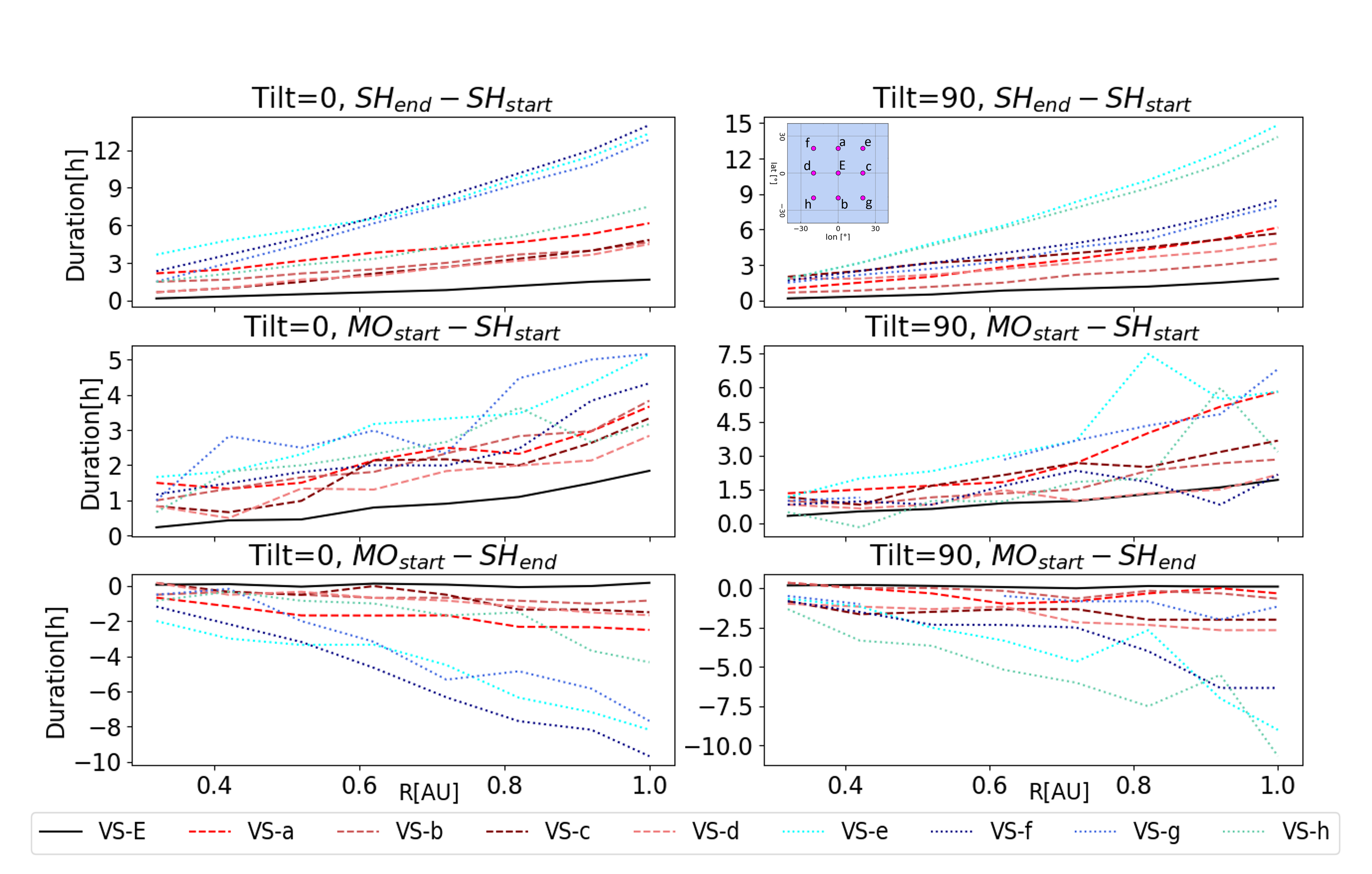}
\caption{Top panels show the extended sheath duration with respect to radial distance for different VS, for both high inclination spheromak (right panels), and low inclination spheromak (left panels). The middle and bottom panels show the same for the clear sheath and MO front region of the MO, respectively. Different colors correspond to different spacecraft: black for VS-E (middle, Earth-directed), red shades for spacecraft on the crosses with respect to VS-E (VS-a, VS-b, VS-c, and VS-d), and blue shades for spacecraft on the diagonals of the spacecraft constellation (VS-e, VS-f, VS-g, and VS-h). For details see main text.
\label{fig7}}
\end{figure}

We first analyse the evolution of the sheath, using different border selections, as explained in Section \ref{boundary_determination}. The top panels of Figure~\ref{fig7} show the duration of the extended sheath in relation to the radial distance from the Sun for the spheromak with low inclination (left panel) and the spheromak with high inclination (right panel). The middle and bottom panels show the same but for a clear sheath and MO front region, respectively. Different colors represent different spacecraft, black for VS-E (middle, Earth-directed), red shades are the spacecraft on the crosses with respect to VS-E (namely the VS-a, VS-b, VS-c, and VS-d), and blue shades are the spacecraft on the diagonals of the spacecraft constellation (namely the VS-e, VS-f, VS-g, and VS-h). For both spheromaks, both clear and extended sheaths increase with increasing radial distance. The largely increasing extended sheath indicates that for most spacecraft, the MO start time is before the sheath end time (similar to what we see for VS-d in the right panels of Figure~\ref{fig2}). This is confirmed by the bottom panels of Figure~\ref{fig7}, where we see that the duration of the MO front region is negative (as per definition $MO_{start}$-$SH_{end}$, where $MO_{start} < SH_{end}$).

The extended sheath shows exponentially increasing profiles with distance for both spheromak inclinations which is consistent with earlier studies based on observations or simulations with other types of FR configurations, such as Gibson-Low (\citealp{Manchester2005}; \citealp{Siscoe2008}; \citealp{Janvier2019}). The study by \cite{Scolini+2021} also confirms this exponentially increasing sheath in the simulation output based on EUHFORIA simulation with spheromak implementation as well as a recent comprehensive observational study by \cite{Larrodera2024}. For the high inclination spheromak (top right panel), we can distinguish three different regimes, based on the rate of change of the sheath. In the black curve (VS-E) it is slowest, i.e. the size of the sheath increases slowly in time in the Earth direction compared to other directions. The rate of change in the four red curves (VS-a, VS-b, VS-c, and VS-d) and two blue-shaded curves  (VS-f and VS-g) is faster compared to the black curve and increasingly faster for each curve (referred to hereafter as the middle increase regime). Finally, the fastest rate of change is for the two curves in the light shades of blue (VS-e and VS-h; referred to hereafter as the high increase regime). This is a direct consequence of the high inclination spheromak northward tilting, i.e. the decrease in $\theta$ angle shown in the uppermost panel of Figure~\ref{fig6}. Due to this tilting, VS-e, and VS -h, which exhibit the high increase regime spend more time in the sheath than the other two spacecraft on the diagonal, VS-f and VS-g. This can also be seen in Figure \ref{fig13} of appendix section \ref{Plasma_Beta}. The situation is slightly different for the low inclination spheromak (top left panel). Again, we see the slowest rate of change for VS-E curve and the middle regime for VS on the crosses (VS-a, VS-b, VS-c, and VS-d). However, the situation is different for the VS on the diagonals of the spacecraft constellation. In this case, VS-h falls in the middle increase regime, while the rest of the spacecraft on the diagonals (VS-e, VS-f, VS-g) are found in the high increase regime. The slight difference in the behavior of the time evolution curves corresponding to the diagonals of the spacecraft constellation is related to the slight differences in the tilting and drift of the spheromak, which is slightly different for low and high inclined spheromak, as well as the inclination itself. Better visualization of the extended sheath crossing for each VS, for low inclination spheromak is shown in Figure \ref{fig14} in appendix section \ref{Plasma_Beta}, while corresponding animations for all the other radial distances are included here: \href{https://doi.org/10.6084/m9.figshare.25849135}{doi:10.6084/m9.figshare.25849135}. Nevertheless, in general, for both inclinations, we observe that the rate of the change in the time evolution curve of the extended sheath is slowest at the apex and increases towards the flanks.

The clear sheath (middle panels of Figure~\ref{fig7}), does not show the regular profile as is the case with the extended sheath in the upper panels. However, the increase in duration with radial distance is still visible. The irregularity of the profile may not be an inherent property of the clear sheath, but rather related to the methodology based on which it was estimated: a) the end and the beginning of the observed interval are determined based on two different methods (the plasma $\beta$ and the magnetic field vector method, see section~\ref{boundary_determination} for details) and b) the time cadence with which the 3D simulation domain data are stored that can lead to an error of $\pm 30$ minutes.

Finally, we analyse the time evolution of the MO front region, which is shown in the bottom panels of Figure~\ref{fig7}. Considering that the duration is here negative (as per definition), we can again see a clear increase in duration with increasing distance from the Sun. Similar to the top panels, the slowest rate of increase is observed for VS-E (black curve), the spacecraft on the cross of the constellation (red color tones) are found in the middle increase regime, and the spacecraft on the diagonals (blue colored tones) are found in the high increase regime. The only difference between low and high-inclination spheromak is seen in the duration of the frontal region for VS-g. For low inclination spheromak, it shows a profile similar to other VS on the diagonals of the constellation, while for the high inclination spheromak, it shows a profile more similar to VS on the crosses of the constellation. This could be related to the slight differences in the tilting and drift of the spheromak for two different inclinations, as noted above for an extended sheath.

\subsubsection{Plasma non-radial flows inside the sheath region}

We next analyze plasma non-radial flows (NRFs) in the extended sheath region, where we refrain our analysis to the extended sheath ($SH_{end}$-$SH_{start}$, see Section \ref{boundary_determination} for details), similarly as was done in the observational study of \citet{Martinic2023}.

\begin{figure}[ht!]
\plotone{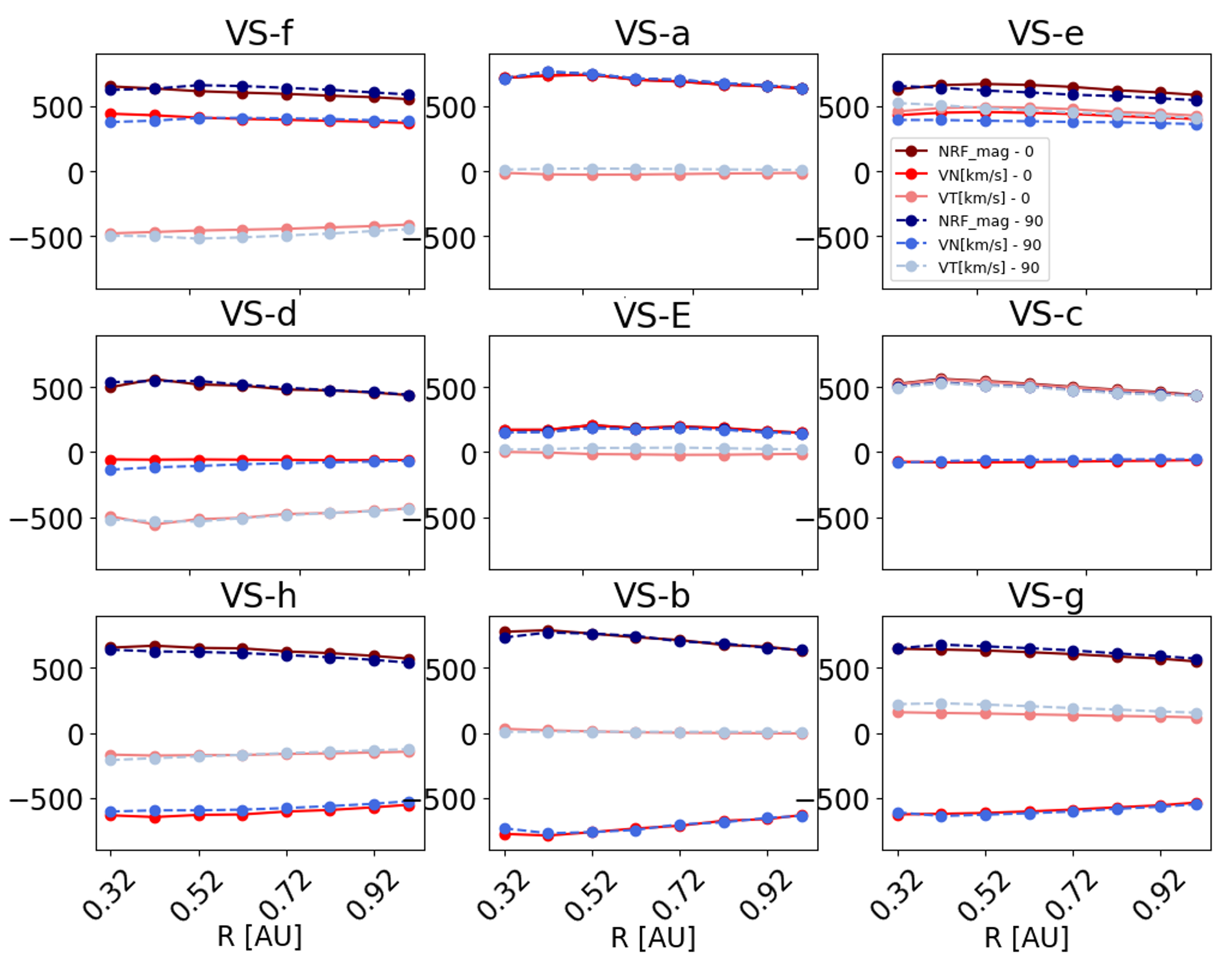}
\caption{The mean value of the NRF magnitude,  the tangential, and the normal component for both low (red shading) and high (blue shading) inclination spheromak in the sheath region in relation to the radial distance.
\label{fig8}}
\end{figure}

\begin{figure}[ht!]
\plotone{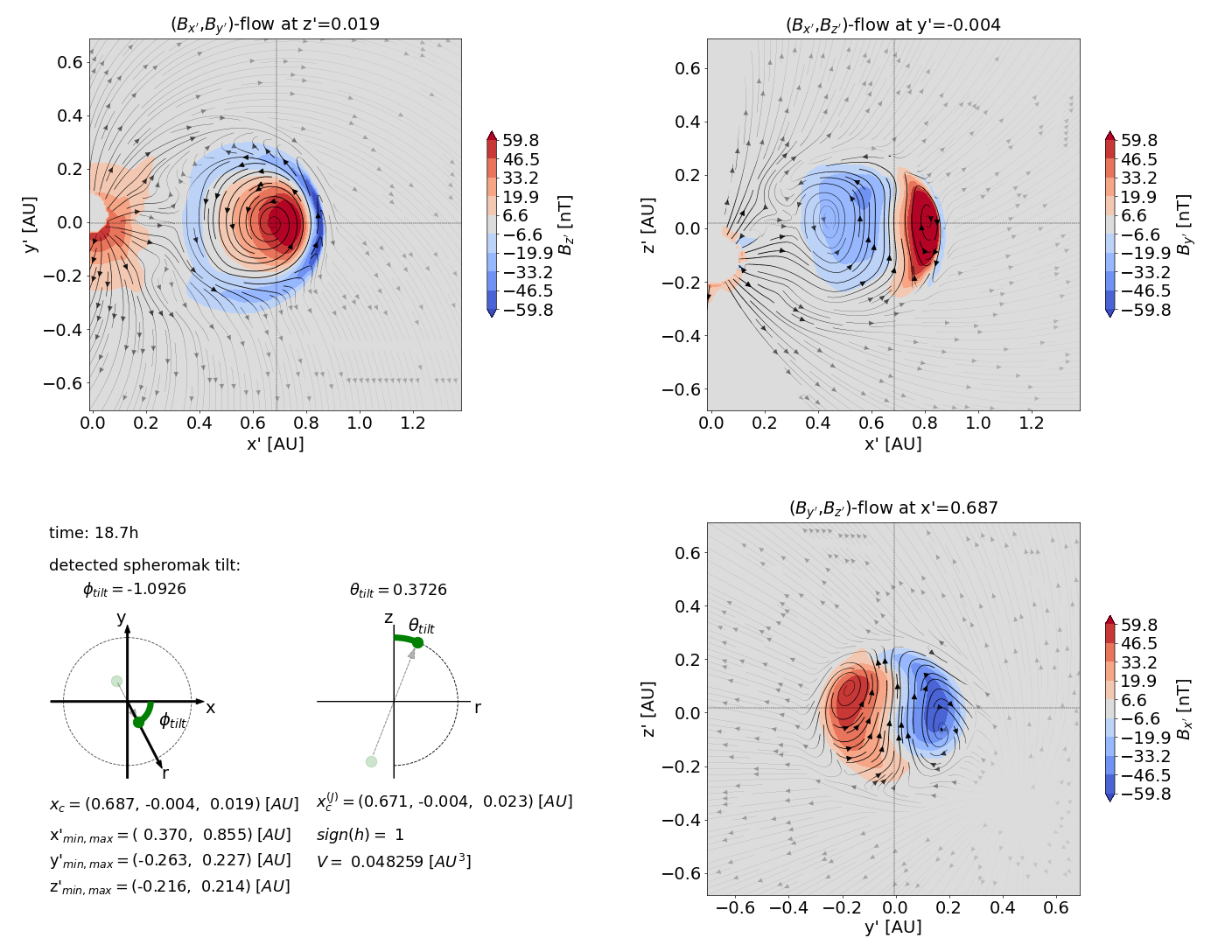}
\centering
\caption{The vector field as distributed inside the spheromak volume in the case of the low inclination spheromak shown in the spheromak's reference frame ($x'$,$y'$,$z'$) for 3 perpendicular planes.
\label{fig9}}
\end{figure}

Figure~\ref{fig8} shows the normal and tangential components, and the magnitude of the NRFs within the extended sheath region. Each subplot represents one VS and the subplots are aligned to mimic the position of the spacecraft on the grid (see Figure~\ref{fig1}). For both high-inclination (blue color tones) and low-inclination (red color tones) spheromak we can see very similar profiles and values of the magnitude of NRFs inside the extended sheath region. This is not surprising considering that a spheromak, regardless of its inclination, is initially a spherical structure. It is important to note that as the simulation progresses they lose their spherical symmetry. More precisely, spheromaks have a perfect spherical structure when they are inserted in the simulation, but as they move through the inner boundary they interact with the ambient IMF and solar wind, and thus, they experience a certain amount of flattening in the plane perpendicular to the central axis. This can be seen in Figure~\ref{fig9} which shows the vector field as it is distributed inside the spheromak volume in the case of the low inclination 18.7~hours since insertion. The detection tool found a spheromak with an extent of 0.490~AU in the East-West direction and 0.430~AU in the North-South direction (see the minimum and maximum values for $y'$ and $z'$ at the bottom left of Figure~\ref{fig9}). These orientations are in the reference frame of the spheromak which is tilted with respect to the HEEQ coordinate frame as given by the angles on the bottom left of the Figure. Due to this distortion of the spheromak shape, for the low inclination spheromak, we have a greater extent in the ecliptic plane compared to the meridional plane, and vice versa for the high inclination spheromak. However, based on the NRF behavior we observe in Figure~\ref{fig8}, it would not seem that this distortion has a significant impact on the NRFs in the sheath region. In the case of the high inclination spheromak, 18.7~hours since insertion, its extent in the East-West and the North-South directions in the spheromak reference frame was 0.500~AU and 0.414~AU, respectively.

The magnitude of the NRFs is lowest for VS-E at 100-200 km/s, which is to be expected as this is the direction of propagation of the two spheromaks, and the solar wind begins to diverge around the MO, while the non-radial flows appear to be more pronounced in the latitudinal and longitudinal directions, away from the apex. Slightly higher values of NRF magnitudes than for VS-E, 400-600 km/s, can be seen for the VS in the ecliptic plane (VS-d and VS-c), while the highest values, 500-800 km/s are seen for VS out of the ecliptic plane. This indicates that the pressure gradient developing in front of the (I)CME acts to slide IMF filed lines away from the center of the (I)CME in the direction perpendicular to the ecliptic plane (see \citealp{McComas1989}). 

\begin{figure}[ht!]
\plotone{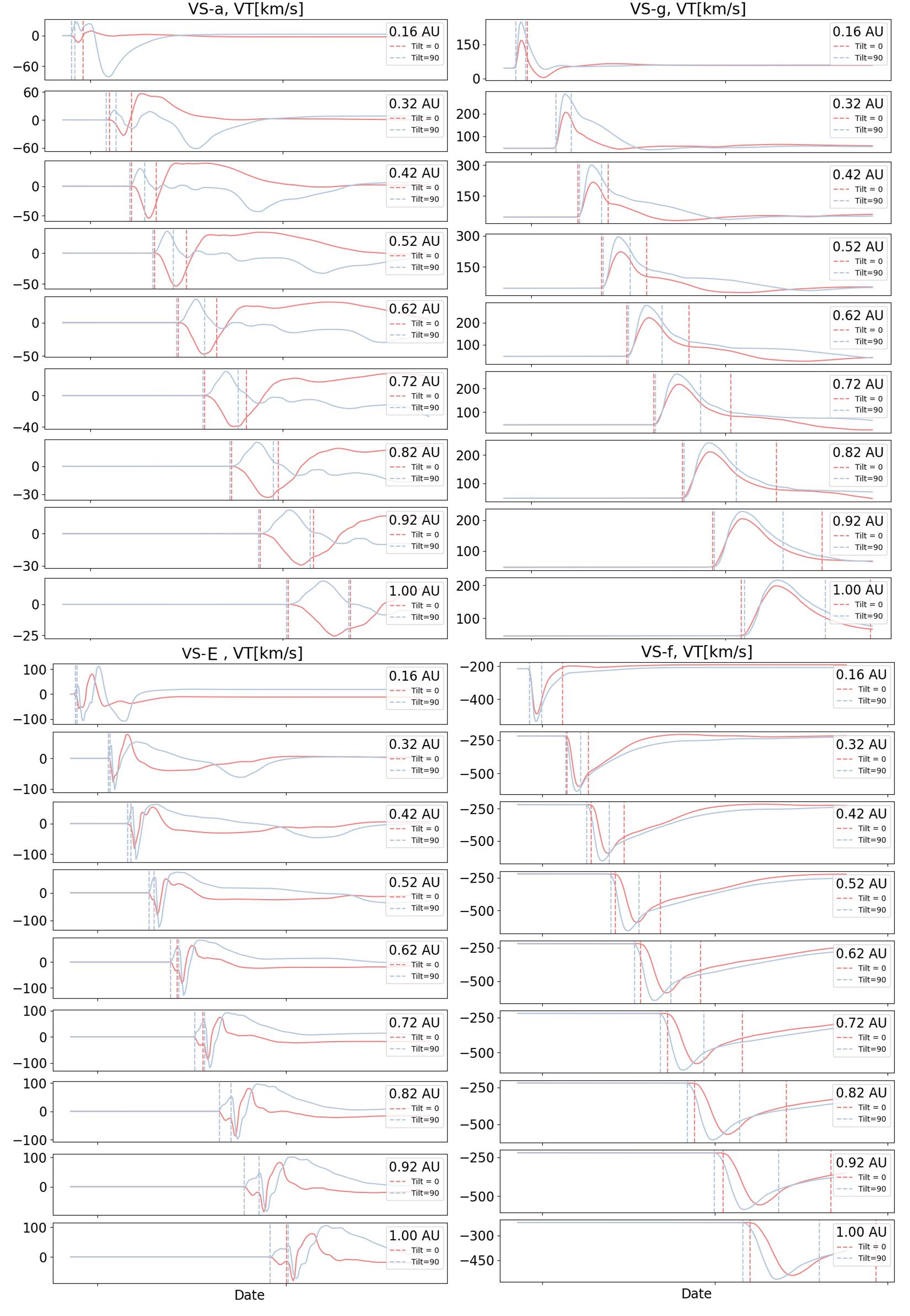}
\caption{The Figure shows the tangential component of the velocity as seen in-situ for SV-a, SV-f, SV-E, and SV-g, from left to right and from top to bottom. The light red lines refer to the low inclination spheromak and the light blue lines are related to the high inclination. Similarly, dashed red vertical lines mark the sheath of the low inclination spheromak, and a pair of light blue vertical lines mark the sheath of the high inclination spheromak.
\label{fig10}}
\end{figure}

For both high and low-inclination spheromak, we see a positive tangential component of velocity for the spacecraft on the right (VS-e, VS-c, and VS-g) and it is negative for the spacecraft on the left (VS-f, VS-d, and VS-h). Similarly, the normal component of the NRFs has positive values for spacecraft above the ecliptic (VS-f, VS-a, and VS-e) and negative for spacecraft below the ecliptic (VS-h, VS-b, and VS-g). As expected, this indicates that the plasma flows from the center of the spheromak toward the edges, bypassing the obstacle.

Some differences between the NRFs for different inclination spheromaks can be observed. First, for VS-f and VS-g, we see a slightly higher tangential component of velocity for high-inclination spheromaks (light blue) than for low-inclination spheromaks (light red). This is consistent with the assumption presented in \cite{Martinic2022} and \cite{Martinic2023} that the ambient plasma bypasses the obstacle in the direction in which the extent of the obstacle is smaller. A spheromak with a high inclination has a slightly smaller extension of the front in the tangential direction than a spheromak with a low inclination, so the tangential component of the NRF is more pronounced in spheromaks with a high inclination. This can also be seen in the right-hand panels of Figure~\ref{fig10}. Figure~\ref{fig10} shows the tangential component of the velocity over all heliospheric distances probed in the simulation for both high (light blue) and low inclination (light red) spheromaks. The right panels refer to VS-f and VS-g, while the left panels refer to spacecraft VS-a and VS-E. We also see a difference in the direction of the tangential NRF component in the sheath for low and high-inclination spheromaks for VS-E and VS-a. The tangential component of the velocity in the sheath has positive values for high inclination spheromaks and negative values for low inclination spheromaks (see Figure~\ref{fig8} center panels and Figure~\ref{fig10} left panels).
Figure~\ref{fig10} shows that the differences in the tangential component of the NRFs for low and high inclination spheromaks are not due to random fluctuations, e.g., due to turbulence, small-scale wave activity, etc., as often seen in in-situ profiles of real (I)CMEs, but rather a consistent difference seen in the simulation in-situ profiles for all heliospheric distances. This difference between the tangential component of NRFs inside the extended sheath region seen in in-situ profiles for VS-E across all radial distances is due to the different drifting (in the east-west) direction of different inclination spheromak as shown in Figure~\ref{fig5}, 4th panel.

The profiles shown in Figure~\ref{fig8} for the NRF magnitude, the tangential, and the normal components show similar behavior, where dominant is the slight decrease with increasing radial distance from the Sun. But we can also see for some VS (for VS on the crosses of the constellation- VS-a, VS-c, VS-b, and VS-d) an NRF maximum is reached somewhere around 0.4~AU, after which a slow decline can be observed. This could be because the rate of growth of the sheath region is higher below 0.4~AU compared to greater distances from the Sun. This is an interesting aspect of the simulation output, but it falls outside the scope of this paper. However, we consider investigating it further in a follow-up study. 

\subsection{The drag parameter}

Finally, we analyse the drag parameter $\gamma$ for two different inclinations. We remind the reader that apart from different inclinations, both spheromaks were inserted with identical properties in the same SW and IMF environment.

\begin{figure}[ht!]
\plotone{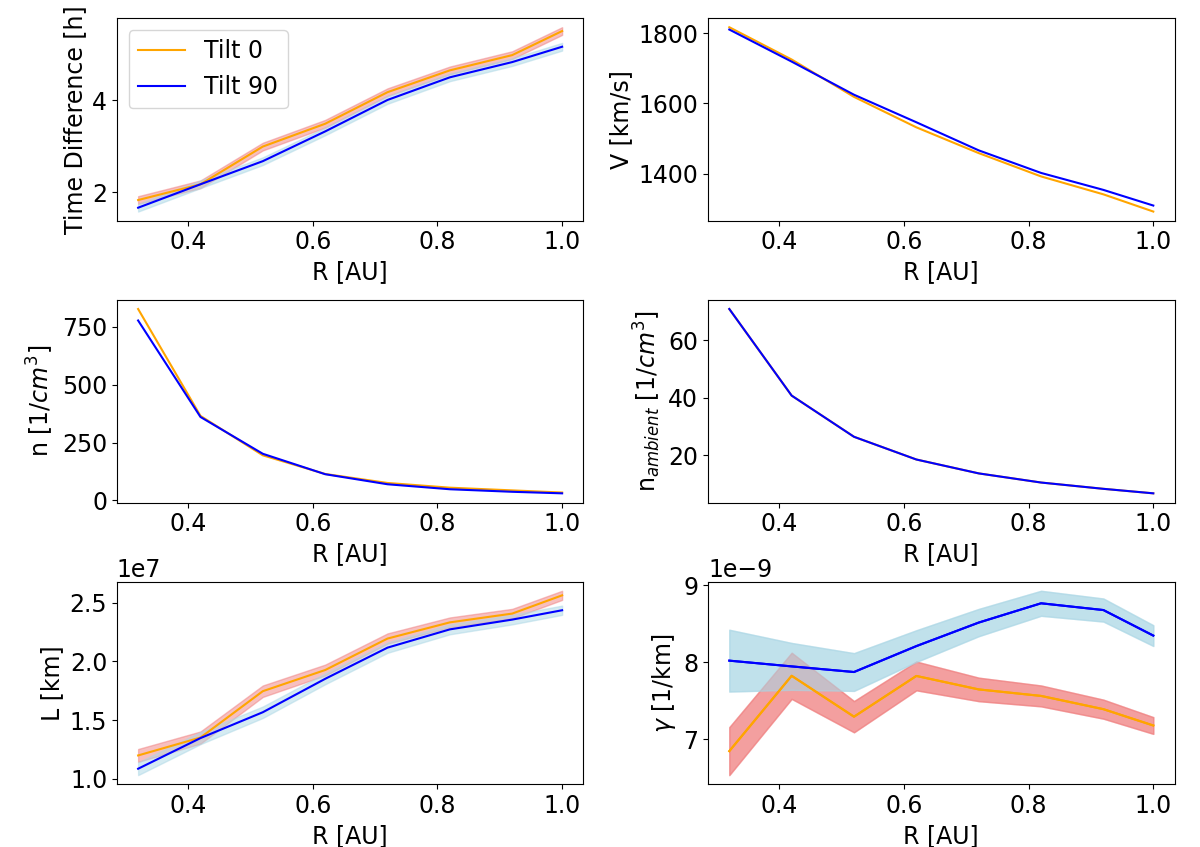}
\caption{The MO duration in seconds based on $B_{vec}$ (corresponding to the green boundaries in Figure~\ref{fig2}), mean velocity and median density inside the respective MO, mean ambient density, radial cross-section of the respective MO, and drag parameter $\gamma$, respectively, from left to right and from top to bottom. Variables are given with respect to radial distance and separately for low inclination spheromak (orange) and high inclination spheromak (blue).
\label{fig11}}
\end{figure}

The bottom right panel of Figure~\ref{fig11} shows the drag parameter $\gamma$ as a function of radial distance for both low (orange) and high (blue) inclination spheromak. The other panels of the Figure show variables needed to obtain the drag parameter 
$\gamma$, based on the formula (2) from section~\ref{Drag_parameter_determination}. From top to bottom and from left to right, Figure~\ref{fig11} shows, 1) the duration of the MO part of the spheromak determined based on the $B_{vec}$ (see green vertical lines shown in Figure~\ref{fig2}, left panels), 2) mean velocity and 3) median density inside the respective MO, 4) the mean ambient density, and 5) the obtained radial cross-section of the MO part. The shaded area shown for the time difference between the MO boundaries (top left panel) takes into account the $\pm$ 5-minute uncertainty in the derivation of the MO boundaries. This uncertainty is also the source of the uncertainties in the calculated values for the radial cross-section $L$ and drag parameter $\gamma$.  We note here that for this portion of the study, we used a higher time cadence for the 3D simulation domain output than what was used for the previous sections. We can see a slight difference between the values of the drag parameter $\gamma$ for low and high inclination spheromak in the simulation. High inclination spheromak has somewhat higher values of the $\gamma$ parameter, meaning that the high inclination spheromak experienced greater drag force in the simulation. We note that this difference is directly related to the difference in the radial cross-section of the MO part (bottom left panel).

From the simulation, it appears that there is a slight difference in the MO duration for low and high inclination spheromak (see top left panel of Figure~\ref{fig11}). The greater the MO duration, the greater the radial cross-section, and then consequently smaller $\gamma$ is obtained. The differences in the mean velocity and median density inside the spheromak MO part (top right panel and middle left panel in Figure~\ref{fig11}, respectively), as well as the differences in the ambient density (middle right panel in Figure\ref{fig11}), are negligible. It is important to note here that the $C_{d}$, dimensionless drag coefficient is taken to be constant and equal to one when calculating the drag parameter $\gamma$ and thus we can not expect the drag parameter $\gamma$ results shown in the bottom right panel of Figure~\ref{fig11} to outline the difference in the draping pattern between the two spheromaks in the simulation. The draping pattern is a crucial factor governing the proposed hypothesis in \cite{Martinic2022} and \cite{Martinic2023} that high inclination spheromak should experience greater drag due to less efficient sliding of the IMF field lines in the direction perpendicular to the motion of the (I)CME \citep[i.e., slipping of the transversal IMF component perpendicular to the direction of (I)CMEs motion, see also][]{Gosling1987}. From Figure~\ref{fig11} we can also see that the drag parameter $\gamma$ is not constant with respect to the radial distance. Except for radial cross-section $L$, drag parameter $\gamma$ also depends on the ratio between the density inside the MO and ambient density, from the middle panels we can see that the decrease in the ambient density shows a slight different profile than the decrease in the spheromak density meaning its ratio changes with radial distance. This result challenges the constant $\gamma$ assumption which is often used in drag-based models in operational mode. Furthermore, the obtained drag parameter $\gamma$ shows a rather low value when compared to for example with results presented in \cite{Vrsnak2013}, where $\gamma$ is in the range 0--2$\cdot$10$^{-7}$. We believe this discrepancy stems from a rather high spheromak density used in our simulation \cite[for a range of densities see also][]{Temmer2021} that increases the density ratio in equation (2) in section \ref{Drag_parameter_determination} and consequently smaller gamma values are obtained.  

When we performed the same analysis of the drag parameter $\gamma$, using equation (2) from section \ref{Drag_parameter_determination}, but with numerical set up with different resolution- $\pm$ 30 min, we found that drag parameter $\gamma$ is slightly higher for low inclination spheromak then for high inclination spheromak (opposite to what is shown in the bottom right panel of Figure~\ref{fig11}). We also cross-checked results show in Figure~\ref{fig11} for drag parameter $\gamma$ using equation (1) from section \ref{Drag_parameter_determination} and we obtained drag parameter $\gamma$ of the same order of magnitude and with similar profile with respect to radila distance, but slightly higher for low inclination spheromak then for high inclination spheromak. Thus, we conclude that different methods and time resolutions showed that the inclination-dependency of drag parameter $\gamma$ cannot be resolved for the given setup.

\section{Conclusions}
\label{Conclusions}

We have investigated the influences of the (I)CME inclination on its propagation in the heliosphere by performing simulations with the EUropean Heliospheric FORecasting Information Asset (EUHFORIA) 3D magnetohydrodynamic (MHD) model \citep{Pomoell2018}. The study focuses on two (I)CMEs implemented as spheromaks \citep{Verbeke2019} with almost identical properties, differing only in their inclination. We also inserted 81 virtual spacecraft into the simulation domain to investigate the radial, longitudinal, and latitudinal differences. One spheromak is embedded as a low inclination spheromak (Tilt=0$^\circ$) with its axis of symmetry perpendicular to the ecliptic and its central axis (toroidal axis of the FR) laying in the ecliptic. The other spheromak is embedded as a high inclination spheromak (Tilt=90$^\circ$), whose axis of symmetry lies in the ecliptic and whose corresponding FR axis is perpendicular to the ecliptic plane. To avoid the CME interaction with heliospheric current sheath and to maximise the inclination effect on the propagation, we have chosen simplified conditions for the solar wind and the IMF. The IMF has a uniform and outwardly directed (positive) radial configuration, $B_{r}=$100~nT at the inner boundary with a radial velocity component of the surrounding solar wind of V$_{r}$ = 400.0~km~s$^{-1}$. Inserted spheromaks were chosen with a high density ($\rho$ = 0.5~$\cdot$~10$^{-17}$~kg~m$^{3}$) and a high velocity (v = 1500.0~km~s$^{-1}$) in order to minimize the spheromak tilting due to torque \citep{Asvestari2022}. Regardless of the fact that we carefully chose the ambient and spheromak conditions, we found that a torque is exerted on the spheromak affecting the orientation and thus, the simulation results should be interpreted accordingly. Furthermore, we derived the boundaries of the spheromak sheath and the magnetic obstacle by using both the in-situ simulation profiles at each VS and determining the plasma $\beta$ boundary as well as the magnetic flux density as a vector field and the magnitude of the plasma pressure gradient derived from the 3D simulation output. We found that the boundaries derived in these two discrete ways are very different.

We found that the duration of the sheath increases exponentially with increasing radial distance, similar to Gibson-Low FR, which is essentially a modified spheromak (\citealp{Manchester2005}; \citealp{Siscoe2008}; \citealp{Janvier2019}). We also found that the duration of the sheath increases towards the flanks of the spheromak, confirming the growth of the sheath as moving away from the (I)CME apex. Similarly, the non-radial flows are smallest near the apex of the (I)CME and increase towards the flanks. However, we found a difference in the magnitude of the non-radial flows in the ecliptic and out of the ecliptic plane. Non-radial flows out of the ecliptic are stronger than in the ecliptic plane for both high and low inclination spheromak. This confirms the assumption of \cite{Martinic2022} and \cite{Martinic2023}, following the discussion in \cite{Gosling1987}, that the transverse IMF magnetic field component "slips" in the direction perpendicular to the (I)CME motion due to the pressure gradient that accumulates in front of the (I)CME during its propagation. We also derived that the drag parameter $\gamma$, a crucial factor in the study of the drag force acting on the (I)CME,  changes slightly with radial distance in the order of $10^{-9}$. We highlighted that such a rather small value of the drag parameter $\gamma$ is a consequence of having inserted very dense spheromaks. 

When cross-checking our results for drag parameter $\gamma$ with the same method, but different resolution, as well as when cross-checking our results with a different method- based on equation (1) from section~\ref{Drag_parameter_determination} we found that inclination dependency of the drag parameter $\gamma$ can not be resolved with our numerical setup. We also note that the drag parameter $\gamma$ determined in this study assumes that the dimensionless drag coefficient $C_{d}$ is constant and equal to one, so our drag parameter $\gamma$ cannot reflect the difference in the draping pattern between the two spheromaks investigated.

Besides the tilting due to torque and drift of the spheromak \citep{Asvestari2022}, the differences between the two differently inclined spheromaks could be masked by the predominantly spherical geometry of the spheromak, which leaves room for progress for the future study in which we plan to use different FR models in the simulation, such as the FRi3D model by \cite{Isavnin2016} among others. In addition, this study showed the potential of MHD simulation studies in linking non-radial flows to the change in the orientation of the surrounding magnetic field (i.e., in revealing the draping pattern), which should also be reflected in the magnetohydrodynamic drag force acting on the CME as it propagates in the surrounding magnetic field and solar wind. In the future, we also hope to explore in more depth the sheath growth rate and the MO front exhibiting both, sheath and MO properties. These topics have been discussed only briefly in this work to keep the manuscript short and with a clear focus.

\begin{acknowledgments}

We gratefully acknowledge the support from the Austrian-Croatian Bilateral Scientific Projects ``Multi-Wavelength Analysis of Solar Rotation Profile'' and ``Analysis of solar eruptive phenomena from cradle to grave''. M.D. and K.M. acknowledge support from the Croatian Science Foundation under the project IP-2020-02-9893 (ICOHOSS). K.M. acknowledges support from the Croatian Science Foundation in the scope of Young Researches Career Development Project Training New Doctoral Students. E.A. acknowledges support from the Academy of Finland/Research Council of Finland (Academy Research Fellow grant number 355659). T.R. is supported by the Swiss National Science Foundation (SNSF) through the grant no.~210064.

\end{acknowledgments}

%






\appendix

\section{Determination of the spheromak boundary using the in-situ profiles}
\label{boundary_determination_apendix}

\begin{figure}[ht!]
\plotone{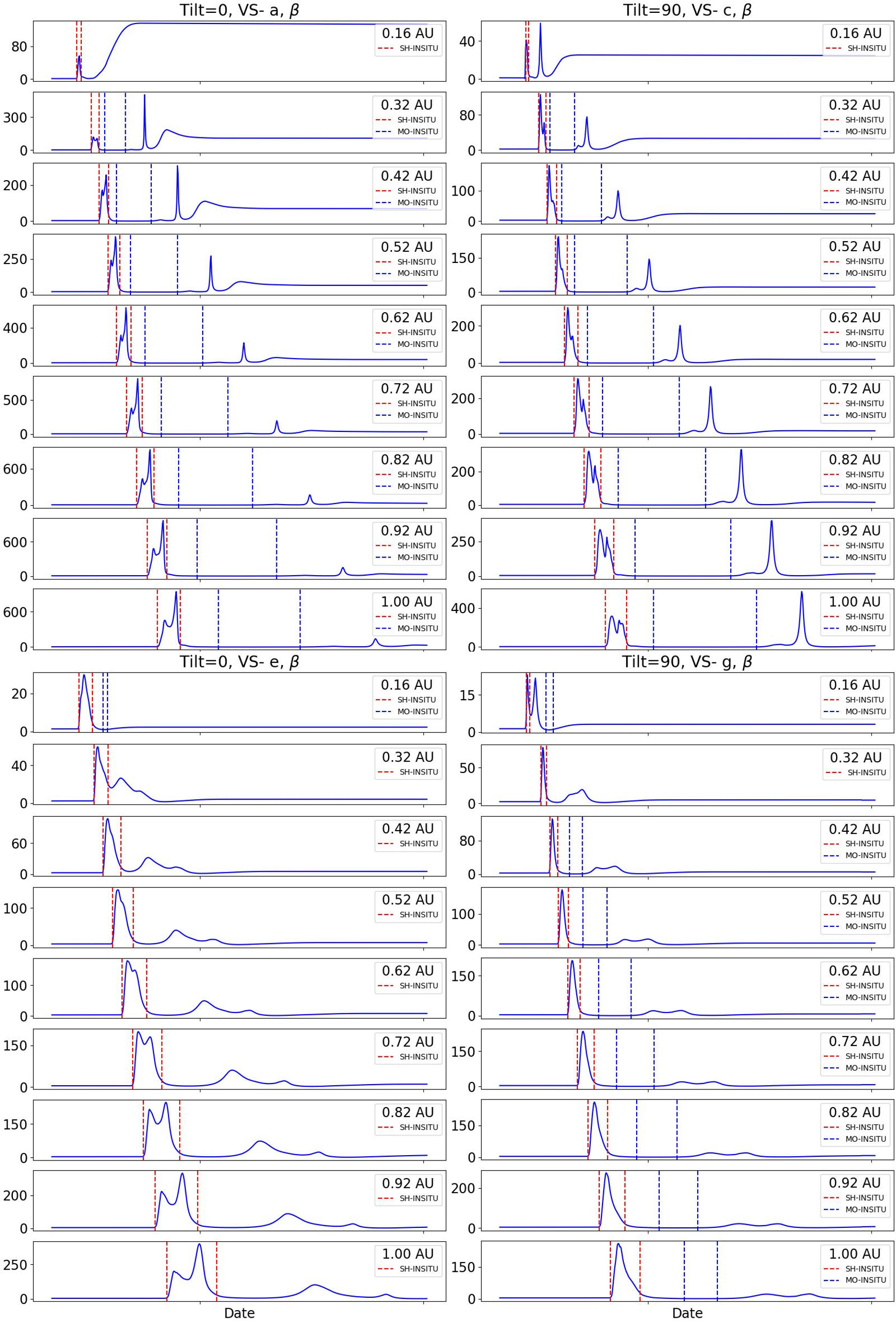}
\caption{Plasma $\beta$ parameter for VS-a and VS-e for low inclination spheromak and for VS-c and VS-g for high inclination spheromak. Red vertical lines mark the extended sheath derived solely based on the plasma $\beta$ parameter increase/decrease, while blue vertical lines mark the MO part of the spheromak where plasma $\beta$ parameter is less than one.
\label{fig12}}
\end{figure}

In Section \ref{boundary_determination} it was already mentioned that the first red vertical line is derived as a 50$\%$ increase of $\beta$ compared to the mean of the first 35 values when iterating from the beginning of the time series. The second red vertical line is derived as a 50$\%$ increase compared to the last 35 values of the time series when iterating from the end to the beginning of the time series. Here we show four exceptions to this method. The second vertical line was derived manually for spheromaks with a low inclination for VS-c and VS-g and spheromaks with a high inclination for VS-a and VS-e. The boundaries for these VS can be seen in Figure~\ref{fig12}. In Figure~\ref{fig12} one can see that the $\beta$ profiles exhibit a numerical artifact, namely an increase in the $\beta$ parameter directly after the MO. Due to this artifact, it was not possible to derive the sheath end boundary using the method described in Section~\ref{boundary_determination}. At this point, we would also like to point out that MO, defined as the part of the in-situ profile where plasma $\beta$ falls below one, is also not visible for all VS and all radial distances. For example, for spheromaks with low inclination, for VS-e (see Figure~\ref{fig12} bottom left panels), there is no plasma $\beta$ less than one for all distances except 0.16~AU, i.e. there is no MO according to the chosen criterion.

\section{Plasma beta parameter spacecraft crossing}
\label{Plasma_Beta}

\begin{figure}[ht!]
\plotone{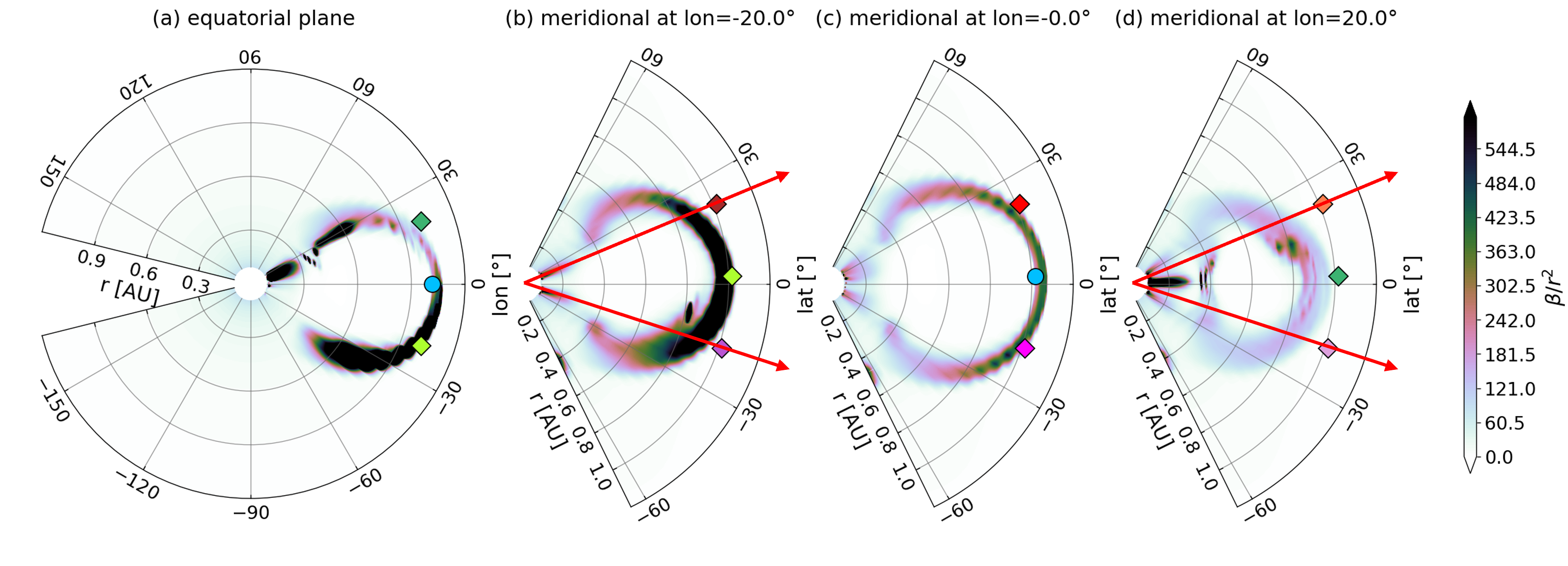}
\caption{ Plasma $\beta$ parameter, normalized by $1/r^{2}$, where r is in AU, of the simulated high inclination spheromak at equatorial and meridional planes.
\label{fig13}}
\end{figure}

\begin{figure}[ht!]
\plotone{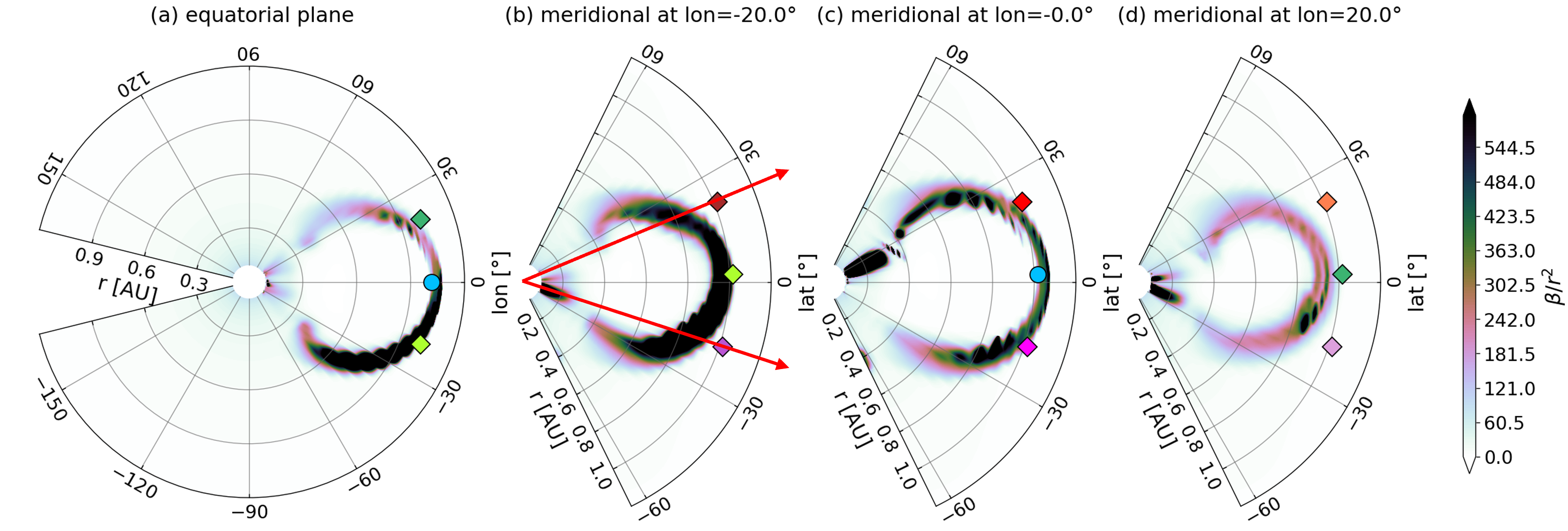}
\caption{Plasma $\beta$ parameter, normalized by $1/r^{2}$, where r is in AU, of the simulated low inclination spheromak at equatorial and meridional planes. 
\label{fig14}}
\end{figure}

Figure \ref{fig13} (for high inclination spheromak) and Figure~\ref{fig14} (for low inclination spheromak) show the plasma $\beta$ parameter, normalized by $1/r^{2}$, where r is in AU, in equatorial and in three meridional planes corresponding to -20$^\circ$, 0$^\circ$, and 20$^\circ$ in longitude, respectively, from left to right. Red arrows indicate the spacecraft crossing. Figure~\ref{fig13} shows spacecraft crossing of VS-f and VS-h in the meridional plane corresponding to -20$^\circ$ longitude and we can see that VS-h will spend more time in the clear sheath region (region of increased plasma $\beta$ parameter) than VS-f. Similarly, VS-e will spend more time in the extended sheath region than VS-g as seen in the meridional plane corresponding to 20$^\circ$ longitude. This confirms the results for the extended sheath duration for high inclination spheromak where we derived the highest increase regime for VS-h and VS-e, while VS-f and VS-g showed a intermediate increase regime (see top right panel of Figure~\ref{fig7}). Analogously, in Figure~\ref{fig14} we can see that VS-h will spend less time in the extended sheath region than VS-f which is why for Vs-h we found a intermediate increase regime of the extended sheath, unlike for the rest of the spacecraft on the diagonal (see top left panel of Figure~\ref{fig7}). 

\bibliography{sample631}{}
\bibliographystyle{aasjournal}



\end{document}